# Machine learning model leveraging SMILES-derived NMR spectroscopy data to predict dopamine D1 receptor antagonists: a prospective framework for forecasting the impact of engineered nanoparticles on the functionalities of small biomolecules

Mariya L. Ivanova[1,*, ORCID], Michael Nicholls [2 ORCID], Nicola Russo[1, ORCID], Gueorgui Mihaylov[3, ORCID], Konstantin Nikolic[1, ORCID]

Author affiliations

[1]School of Computing and Engineering, University of West London, London, UK

[2] The University of Law , UK

[3] Haleon, UK

*Corresponding author 21435612@student.uwl.ac.uk

## Abstract

The article proposes a conceptual approach for evaluating the impact of engineered nanoparticles (NPs) on the functionality of small biomolecules. The developed machine learning (ML) model is based on in-silico 13C NMR spectroscopy chemical shifts derived by the SMILES notations on small biomolecules. The rationale behind this approach is that 13C NMR provide information about the atom environment of the carbon atoms. Thus, decomposing the small biomolecules into their fundamental 13C NMR spectral data, and performing classification based on the count and position of chemical peaks, establishes a baseline for evaluating the impact of NPs on the functionality of small biomolecules, even if the ML model is not based on nano data. The approach mitigates not only the scarcity of nano-bio data but also hold potential for building of NP`s portfolio by utilising data collected from various *in vitro*, *in situ*, *in vivo*, and organ-on-a-chip environments across multiple timeframes. Such a framework enables predictive modeling based on these multi-environmental datasets, facilitating a deeper understanding of NP behaviour. The methodology was demonstrated using data from bioassay focused on human dopamine D1 receptor antagonists provided by PubChem. The model was train with 26,766 samples and test on 5,466 samples, achieving Accuracy of 70.8%, Precision of 74.3%, recall of 63.6%, F1-score of 68.5% and ROC of 70.8% were achieved by the Support Vector classifier, with an Area Under the Curve (AUC) of 76% and Matthews Correlation Coefficient, MCC=0.4204. A secondary, non-NP-related ML model was developed to complement the study case. It uses PubChem compound and substance identifiers (CIDs and SIDs) to predict whether pre-designed small biomolecules have the potential to be human dopamine D1 receptor antagonists.

## 1. Introduction

The challenges facing accurate nanoparticle (NP) prediction are multifaceted, primarily stemming from the inherent complexity of the materials and the data limitations in the field [1]. Predicting NP behaviour, whether stability, toxicity, or biological efficacy, requires modelling the high-dimensional interplay of numerous variables, including size, shape, surface chemistry, and core material. A major obstacle is the data bottleneck; experimental datasets are often small, sparse, and suffer from a lack of standardisation across different laboratories, making it difficult to train robust and generalisable machine learning (ML) models. Furthermore, predicting the nano-bio interface is critically challenging, as the immediate formation of a protein corona upon entering a biological environment fundamentally alters the NP's identity and subsequent fate, creating a complex, dynamic system that computational models struggle to fully capture. This combination of complex physics, inconsistent experimental reporting, and dynamic biological interactions creates a significant translational gap between *in vitro* predictions and *in vivo* outcomes.

An innovative approach is adopted in this study to mitigate the scarcity of nano-bio interaction data. While Nuclear Magnetic Resonance (NMR) spectroscopy is generally not employed for the direct analysis of NPs due to the solid and inaccessible nature of the NP core, 13C NMR spectrometry is utilised here as a strategic alternative. This method provides high-resolution data regarding the electronic environment of the carbon atoms within the small biomolecules, serving as a proxy for the influence of the nanoparticle interaction [2]. Since the carbon skeleton forms the structural basis of small biomolecules, and their functionality is intrinsically linked to this skeleton, it was hypothesised that functionality could be predicted based on 13C NMR chemical shifts. This approach is deemed applicable to the presence of NPs because the atomic environment of the carbon skeleton is shifted by NP interactions. Rather than attempting to analyse the inaccessible NP core, focus is directed toward the functional impact of the NP on the attached compound. The 13C NMR chemical shifts were used as a sensitive molecular proxy. These shifts are regarded as acutely responsive to changes in the electronic environment of the drug; therefore, perturbations caused by NP attachment may be detected and directly linked to biochemical activity, pKa changes, and hydrogen bonding. Based on these high-fidelity data, an ML model was developed, which, despite not being trained on physical NP parameters, holds the potential to forecast the ultimate influence of NPs on biomolecular functionality.

The 13C NMR spectroscopy utilises the magnetic property of the 13C isotope [3], measuring the atomic nuclei's absorption and emission of radiofrequency radiation. The 13C nucleus contains six protons and seven neutrons. The extra ½ spin makes the 13C isotope behave as a little magnet. As a result, NMR spectroscopy is sensitive to differences in the chemical structure of the organic molecule, giving information for the connectivity and relative position of the carbon atoms in the molecule. So, given the Carbon skeletons of the organic molecule and the technology of 13C NMR spectroscopy, it was hypothesised that 13C NMR spectroscopic data would contribute to the development of robust, reliable, cost- and time-effective ML model that would predict both the functionalities of small biomolecules and the impact of NPs on these functionalities [4].

Various studies have been conducted exploring the combination of ML and 13C NMR spectroscopy, such as the study of correlations between peak patterns and chemical structure using machine learning [5]; a computational approach to establish correspondence between molecular graphs and NMR spectra [6]; prediction of 13C NMR chemical shifts with a message passing neural network (MPNN) [7]; detection of chemical shifts of benzene compounds [8]; software predicting 13C spectroscopic data by uploading SMILES, international chemical identifier (inChI) or drawing the structure of the compound [9]; prediction of NMR spectra with the COLMARppm software using NMR data obtained in aqueous solution [10]; use of an artificial neural network (ANN) to generate new structures based on information obtained from 13C NMR [11]. On the other hand, ML has been increasingly integrating into the search for innovative, robust, and reliable approaches that can significantly reduce the time and cost of laboratory experiments and, consequently, the time and cost of the research itself. Some examples of such approaches are AlphaFold3 for modelling the structure of proteins and their interactions [12]; deep learning for structure-based drug design [13]; generation of ML attributes based on atomic characteristics [14]; machine learning approaches for pharmaceutical decision-making processes [15]. Moreover, ML applications regarding the dopamine D1 receptor antagonists are classification conducted using topological descriptors [16]; effective identification of novel agonists of dopamine receptors has been obtained using predictive or generative geometric ML [17]; agonists and antagonists have been predicted, using topological fragment spectra (TFS)-based support vector machine (SVM) [18]. However, the approach presented in the current article has not been reported so far in the available literature.

Data used for demonstrating the hypothesis was provided by PubChem, the world's biggest freely available database for chemical data. It was chosen the PubChem AID 504652 bioassay focused on selecting and developing allosteric modulators of D1 DAR for in vivo and in vitro applications [19]. It was performed by measuring the compound D1 DAR antagonism with an EC80 addition of dopamine by tracking calcium flux in a force-coupled inducible Hek293 Trex D1 cell line. The compound concentration utilised in this bioassay was 10.0microM. For more information regarding the screening protocol, please refer to the bioassay`s documentation [19]. Following the bioassay, SMILES notations were extracted from the dataset and converted into predicted 13C NMR spectroscopic data using the NMRDB software [20]. This publicly available online tool utilizes Hierarchically Ordered Spherical Description of Environment (HOSE) codes for chemical shift prediction, with a reported mean error of approximately 1.2 to 2.8 ppm. It must be emphasised that the prediction of nanoformulations (NFs) currently results in significantly higher error margins; consequently, predicted data for NFs were excluded from the generation of in silico datasets. To ensure maximum accuracy and reliability, the 13C NMR spectroscopic shifts for all tested nanoformulations must be obtained using a physical 13C NMR spectrometer rather than computational estimation.

To complement the primary investigation, a secondary ML model, called a CID_SID ML model, was developed utilising the same bioassay data established for the main study. The name of this secondary ML model derived from the implementation of PubChem Compound (CID) and Substance (SID) identifiers, which are leveraged to predict the latent functional outcomes of small molecules beyond their initial therapeutic design [21]. The CID_SID ML model is not nano-bio related study.

The rationale for the CID_SID ML methodology is based on the nature of PubChem’s indexing system. CIDs are generated based on the unique chemical structure, representing a standardised, non-redundant molecular identity [22]. SIDs, conversely, provide a broader layer of metadata, encompassing source-specific information, experimental context, and chemical variations as provided by different depositors. By integrating both identifiers, the model captures a comprehensive profile of the small biomolecule's chemical and historical bioactivity landscape. While existing literature has documented the application of similar CID_SID ML models across diverse case studies, it is specifically positioned as a building unit and argued that the integration of such models into development pipelines could significantly enhance risk assessment protocols during clinical trials. By relying solely on PubChem identifiers, this approach offers a computationally and time-efficient methodology for identifying the latent potential or secondary effects of drug candidates beyond their initial design.

## 2. Methodology

### 2.1. Main study methodology

In the field of nanomedicine, terminology is dictated by the functional roles assigned to each component, typically categorising the larger entity as the carrier and the smaller as the payload or ligand. When nanoparticles accompany small biomolecules, the biomolecule serves as the primary targeting ligand, such as folic acid or a specific peptide, that guides the NP to a receptor, effectively making the NP a passenger that provides secondary functions like imaging contrast. Conversely, the more standard framework involves small biomolecules accompanying NPs, wherein the NP acts as a protective vehicle or carrier for its molecular cargo. In this context, the NP improves the bioavailability and solubility of the small-molecule inhibitors by shielding them from premature degradation within the body.

In the presented study, however, the traditional hierarchy of components is not considered; instead, the functional impact of the nanoparticle on the biological activity of the small molecule is prioritised, regardless of whether the nanoparticle serves as the carrier or the payload. The conjugation between the nanoparticle and the biomolecule is treated as an established premise, and the optimisation of the delivery purpose of this conjugation is excluded from the scope of the investigation. Rather, the research is focused on modeling how the presence of the nanoparticle influences the molecule over time across various environments.

As noted above, the study utilized quantitative high-throughput screening (qHTS) data targeting human dopamine receptor D1 (D1 DAR) antagonists, sourced from the PubChem AID 504652 bioassay [19]. While the raw dataset comprised 359,035 samples and 10 feature columns, the analysis focused on a specific subset of attributes, such as PubChem compound identifiers (CID), SMILES notations, and primary screening outcomes. The sequential workflow for processing this data is illustrated in Figure 1.

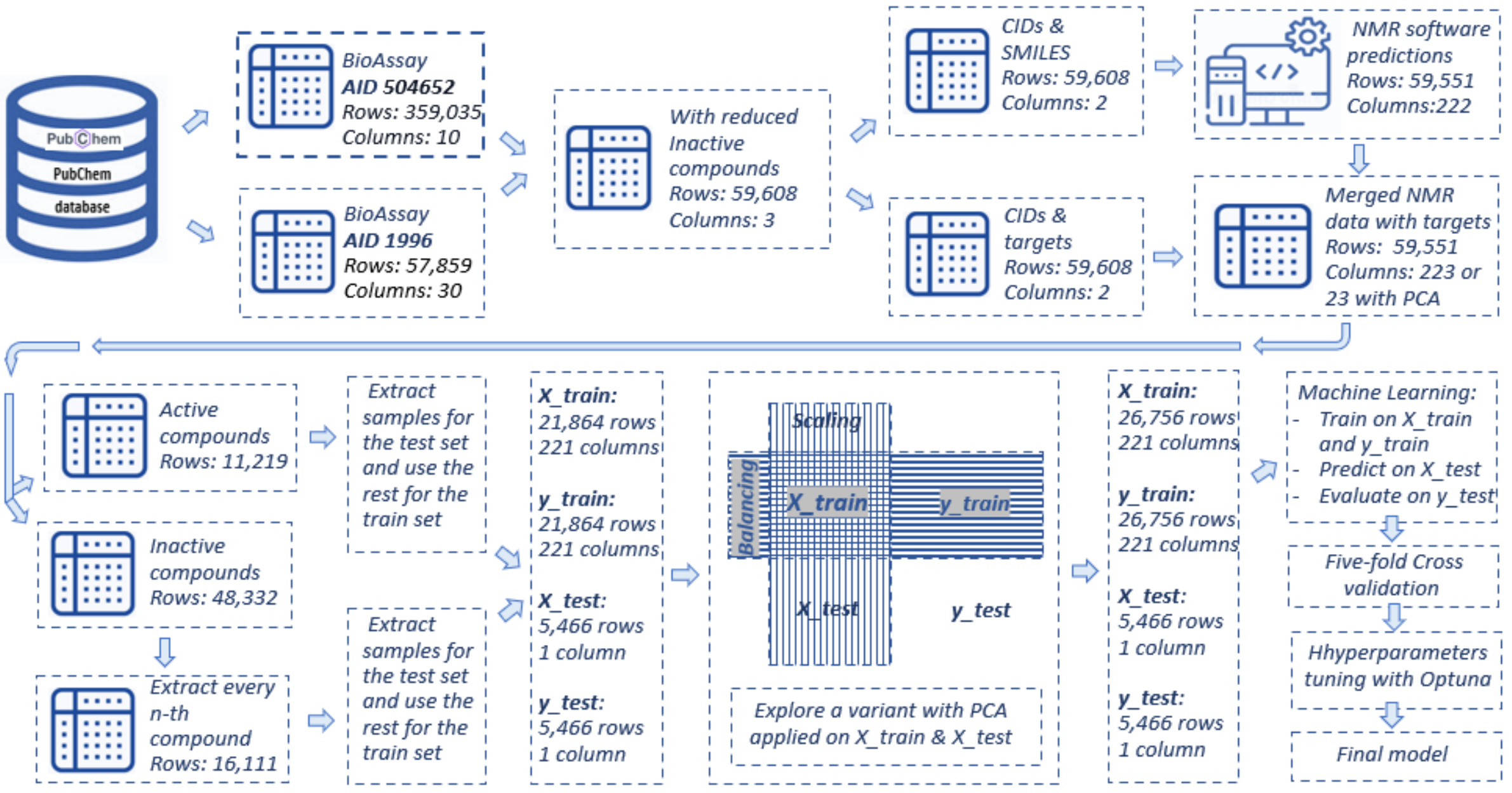

Figure 1.
$^{13}$C NMR spectroscopy data-based ML model
for predicting of human dopamine D1receptor antagonists' methodology

To address the severe class imbalance within the PubChem AID 504652 bioassay, initially consisting of 347,569 inactive vs. 11,466 active compounds, a strategic data-reduction protocol was implemented. The PubChem AID 1996 solubility bioassay was utilised as a filter to reduce the inactive small biomolecules [23]. By merging both bioassays and retaining only common compounds based on CID and SMILES notations, the inactive set was narrowed to focus on a chemical space characterized by reliable physicochemical properties. Although a selection bias was introduced by this process, the shift was intentionally designed to prioritize molecules relevant to drug discovery.

Importantly, the aqueous solubility labels from PubChem AID 1996 bioassay were disregarded; the dataset was used strictly as a mechanism for reducing the number of inactive D1 receptor antagonists. To ensure the model remained focused on its primary objective, detecting active D1 antagonists, any active compounds excluded during the merge were subsequently restored and combined with the reduced inactive pool. Through this dual-action strategy, both the extreme data imbalance and the selection bias were successfully mitigated. The resulting dataset comprised 59,608 total samples, containing all 11,226 active compounds and the remaining 48,382 inactive samples.

From the curated collection, two specialised datasets were derived for subsequent analysis. The first, designated as "smiles", consisted of the CIDs and SMILES notations and was utilized to generate 13C NMR spectrum chemical shifts. The second, labeled "targets", was created to store the CIDs alongside their corresponding screening outputs. At the conclusion of the data generation process, the 13C NMR spectroscopic features were merged with the "targets" dataset. This final integration produced a comprehensive dataset for machine learning, pairing structural spectroscopic data with the definitive labels for dopamine D1 receptor antagonism.

To automate the acquisition of predicted 13C NMR spectroscopic chemical shifts, the Selenium web-scraping framework was employed to interface with the publicly accessible nmrdb.org platform [20, 24]. For each sample, the SMILES notation was programmatically uploaded to the server via a generated text file. Following the platform's generation of a graphical molecule representation and a corresponding table of chemical shift values, the number of peaks was extracted and categorized into specific subranges defined by natural numbers. A structured data frame was then constructed, in which the features were defined by these spectral subranges and the cell values represented the frequency of chemical shifts observed within each range. Please, see the illustration of a samples in Fig 2. To complete the dataset for ML, biological activity labels from the PubChem AID 504652 bioassay were integrated as target variables [19]. Finally, the feature space was refined by removing columns containing only zero values, resulting in a reduction of the spectral subrange columns to 224. During this transformation process, the initial dataset of 59,608 samples was reduced to 59,565. This minor loss of data occurred because the NMRDB tool was unable to generate spectroscopic predictions for a small subset of the submitted molecular structures. It was necessitated by the inherent algorithmic constraints of the NMRDB prediction engine [20]. These failures typically occur when a molecular structure contains rare chemical fragments that are not represented in the underlying HOSE (Hierarchical Organization of Spherical Environments) code database, preventing the software from performing an accurate fragment-based lookup. Additionally, structural discrepancies, such as non-standard SMILES syntax, hypervalent atoms, or complex stereochemical configurations, can exceed the parser's recognition parameters. Consequently, these 43 samples were excluded to ensure that the resulting ML model was trained exclusively on high-confidence spectroscopic descriptors, thereby maintaining the integrity and consistency of the feature set.

To further refine the class balance, the inactive compound population was subjected to a second reduction phase. A randomised systematic 1:3 downsampling approach was implemented, in which every third compound from the randomised inactive pool was retained. These samples were then concatenated with the full set of active compounds from the PubChem AID 504652 bioassay, resulting in a final dataset of 27,330 samples. This specific downsampling strategy was preferred over standard automated undersampling techniques to ensure the structural integrity of the chemical landscape was preserved. While standard random undersampling is often susceptible to stochastic aggregation, which can over-represent certain chemical clusters while leaving gaps in others, the application of a uniform stride following an initial random shuffle guaranteed an even distribution of samples. Furthermore, this method avoided the black-box nature of complex algorithmic undersampling, such as NearMiss or Tomek Links, which may inadvertently distort decision boundaries or introduce selection bias by selectively removing data points based on proximity [25] . By utilising a deterministic mathematical rule, absolute reproducibility was ensured across different computational environments. Ultimately, this systematic reduction lowered the data resolution to a scale that was computationally manageable for high-dimensional 13C NMR analysis without altering the underlying density or statistical shape of the inactive population.

From the finalised dataset of 27,330 samples (comprising 11,226 active and 16,104 inactive D1 receptor antagonists), a balanced testing set was partitioned. By extracting 2,733 samples from each class, a test set representing 20% of the total volume was established. This procedure was implemented to ensure metric reliability during evaluation and to circumvent classification bias, where imbalanced data can produce deceptively high accuracy by favouring the majority class. Within this balanced framework, the accuracy score provides a direct, uninflated reflection of the model’s true generalisation capability.

Furthermore, the convergence of micro-averaged and macro-averaged metrics, specifically Precision, Recall, and F1-score, simplifies the analysis, allowing the Macro F1-score to serve as a robust indicator of performance across both classes. This approach ensures that the model is penalised equally for misclassifications in either the active or inactive populations, providing a transparent assessment of its ability to discriminate between functional outcomes. The remaining 21,864 samples were allocated to the training set; to rectify the internal distribution, the Random Over Sampler algorithm was applied, resulting in a finalised training sets of 26,756 samples [25].

The ML binary classification estimators utilised in this study were sourced from the Scikit-learn library [26]. A comparative analysis was conducted among several algorithms, including K-Nearest Neighbours (KNN), Random Forest Classifier (RFC), XGBoost Classifier (XGBC), Decision Tree Classifier (DTC), Gradient Boosting Classifier (GBC), and Support Vector Classifier (SVC), to identify the optimal estimator for this investigation [25, 27-31]. To prevent the arbitrary selection of a model, the top-performing candidates were initially evaluated based on their statistical significance between each other. Subsequently, five-fold cross-validation was performed on the selected classifier to rigorously confirm its generalisability and ensure consistent performance across diverse data partitions.

To address the high dimensionality of the dataset, stemming from the conversion of SMILES notations into spectral data and the generation of features corresponding to natural-number subranges of 13C NMR chemical shifts, Principal Component Analysis (PCA) was employed [32, 33]. The optimal number of components for the dataset was determined through an unsupervised ML approach; following this calculation, the suggested number of components was applied to execute the dimensionality reduction, ensuring that the most significant variance was preserved while minimising computational redundancy.

The metrics suitable for binary classification used to evaluate the ML models were as follows [26]:

(i) Accuracy was calculated by dividing the correct prediction by the total number of predictions, unveiling the overall correctness of the ML model.
(ii) Precision exposes how many of the positive predictions are true positive
(iii) Recall reveals how many of the positive cases in the data set have been identified correctly
(iv) F1-score is the harmonic mean of the precision and the recall
(v) ROC curves illustrate how a binary classifier's performance changes as its decision threshold for classifying instances varies.

Hyperparameter optimisation was conducted using Optuna, a sophisticated Bayesian-based framework designed for automated search space exploration [34]. This approach utilises the Tree-structured Parzen Estimator (TPE) algorithm to iteratively refine the selection of parameters by balancing exploration of the search space with the exploitation of previously identified high-performing regions.

The hyperparameters integrated into this tuning process, along with their respective ranges and configurations, were as follows:

(i) The ‘rbf’ kernel (radial basis function kernel) finds boundaries and achieves high accuracy by mapping data into a higher-dimensional space, separating non-linearly separable data; the kernel ‘linear’ - the choice of the SVM when the data can be effectively separated by a linear boundary.

(ii) The parameter C minimises the training error, controlling the balance between the closest data points and the separating hyperplane.

(iii) Gamma, which controls the influence of a single training example

Prior to defining the final hyperparameter configuration, the machine learning model was scrutinised for overfitting by monitoring the divergence between training and testing accuracy. A maximum deviation threshold of 5% was established, a value selected to accommodate the high-dimensional nature of the 13C NMR data while maintaining the robust generalisation required for effective virtual screening. This specific margin was deemed sufficiently narrow to ensure that the model did not merely memorize stochastic noise, yet lenient enough to permit the capture of the complex, non-linear relationships inherent in 13C NMR chemical shifts.

The identification of this early stopping criterion served as the terminal validation step before the definitive model was selected. Subsequently, model performance was visualized through the generation of learning curves, confusion matrices, and Area Under the Curve (AUC) plots, alongside a comprehensive classification report. Furthermore, the true predictive power was elucidated via the Matthews Correlation Coefficient (MCC)[35]; this provided a balanced evaluation of the model's discriminative ability, effectively revealing performance nuances that standard metrics, such as Accuracy or the F1-score, frequently obscure.

## 2.2. Complementary study methodology

The methodology for the complementary CID_SID ML model was adapted from the original study, the workflow of which is visualised in Figure 2 [ 21, 36-39] . A dataset focused on dopamine D1 receptor antagonists was generated using a procedure similar to the one described above, with inactive compounds reduced and test sets partitioned in an identical manner. Unlike the primary study, this dataset was restricted to three columns: CID, SID, and labels. However, since NMR spectroscopy transformation was not performed for the CID_SID ML model, all samples were preserved, resulting in a final dataset of 27,665 small biomolecules. To generate the test sets, the shuffled dataset was partitioned by randomly selecting 2,767 samples per class, culminating in a 20% test set (totalling 5,534 samples). The remaining data was used for training; following a balancing procedure, the training set consisted of 22,131 samples. Thus, the CID_SID ML model was trained on 22,131 samples and evaluated against a test set of 5,534 samples

While the established ML learning workflow was followed, two additional models were developed using MORGAN2 fingerprints and RDKit SMILES transformed data [40, 41]; these were utilised to facilitate a direct performance comparison with the CID_SID ML model. Although Graph Neural Networks (GNNs) currently represent the state-of-the-art in molecular modeling, they were excluded from this comparative analysis [42]. The primary objective of the CID_SID ML model extends beyond raw predictive accuracy to encompass computational efficiency and processing speed, benchmarks that GNN architectures, due to their high resource demands, often fail to meet.

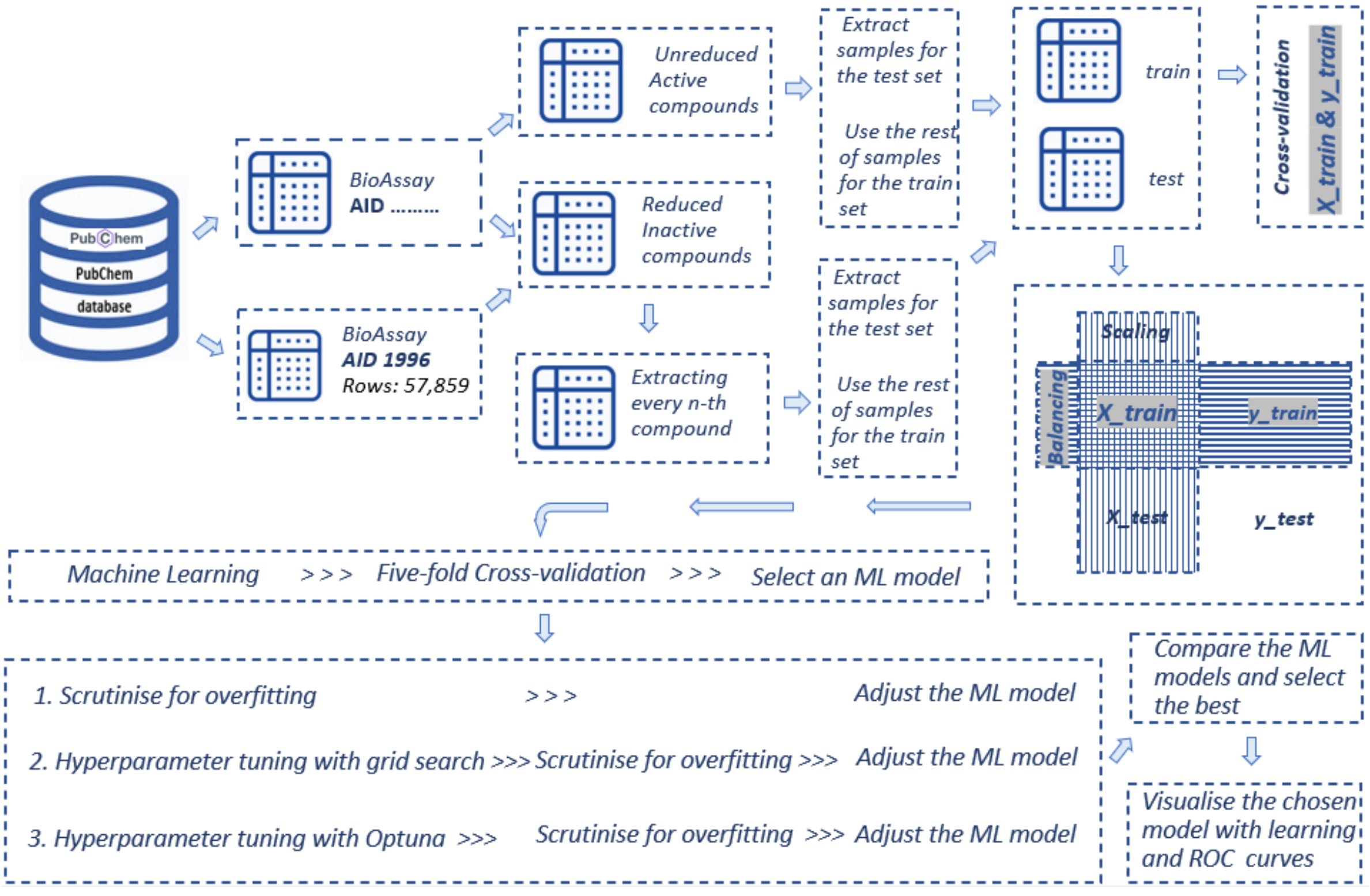

Figure 2. CID_SID.
Methodology for development of a CID-SID ML model
(source Ivanova et al., 2025 [21])

## 3. Results

### 3.1. Main study's results

Initially, from the above-listed ML classifiers, SVC was the optimal variant, achieving Accuracy of 71.4%, Precision of 77.1%, Recall of 61.0%, F1-score of 68.1% F1 and ROC of 71.4%, followed by XGBC with Accuracy of 69.4%, Precision of 73.6%, Recall of 60.4 %, F1-score of 66.3% and ROC of 69.4%. A comparative analysis of the performance of these two ML models revealed that the difference between the SVC and the XGBC is statistically significant. Specifically, SVC demonstrated a performance metrics advantage that significantly outpaced XGBC, suggesting it is the superior model for this specific dataset. The mean values of the metrics obtained by five-fold cross-validation for the SVC and the considered dataset were Accuracy of 0.7428±0.0024, Precision 0.6875±0.0047, Recall of 0.6183±0.0082, F1-score of 0.6510±0.0043 and ROC of 0.7796±0.0036. Superior performance was demonstrated by the cross-validation model, which achieved a higher Accuracy of 0.7428±0.0024 and a significantly stronger ROC-AUC of 0.7796±0.0036, compared to the 70.8% Accuracy and 70.8% ROC recorded in the single iteration. Although a higher Precision of 76.4% was produced by the single SVC run, it was accompanied by lower scores in Recall and overall discriminative power, suggesting a less balanced classification approach. Furthermore, high statistical reliability was evidenced by the narrow standard deviations observed across all cross-validation folds, confirming that a more stable and generalised performance is ensured by the ensemble approach than by a singular data split. While individual runs may yield isolated peaks in specific metrics like Precision, the cross-validation framework provides a more authentic and stable representation of model performance. The superior ROC-AUC and consistent accuracy scores achieved through iterative testing confirm that the models are well-calibrated and capable of generalising to unseen data. By prioritising these robust, multi-fold results over single snapshots, a more reliable foundation is established for the model's deployment in complex decision-making environments.

To evaluate the model's generalisability and guard against overfitting, the performance of the SVC was scrutinised across a range of regularisation parameters C using a 5% maximum divergence threshold between training and test balanced accuracy (Figure 3) . At C = 0.001, the model exhibited significant underfitting, yielding a balanced accuracy of only 0.500 for both sets. As C increased, performance improved; however, the model remained within the established generalisation threshold only up to C = 0.150 (Train: 0.738, Test: 0.691, Gap: 4.7%) (Figure 4). However, more fine tuning revealed that at C=0.190 the ML model was not overfitted yet.  Beyond this point, specifically at C = 0.200, the divergence reached 5.9%, indicating the onset of overfitting. At higher values (C>100), the training accuracy approached 1.000 while test performance stagnated or declined, clearly demonstrating the variance-bias trade-off. Consequently, C values between 0.10 and 0.20 were identified as the optimal range for maintaining a robust balance between predictive power and generalisability for the small biomolecule dataset. So, the ML model with default hyperparameters and C=0.19 achieved Accuracy of 69.6%, Precision of 71.3%, Recall of 65.7%, F1-score of 68.4% F1 and ROC of 69.6%

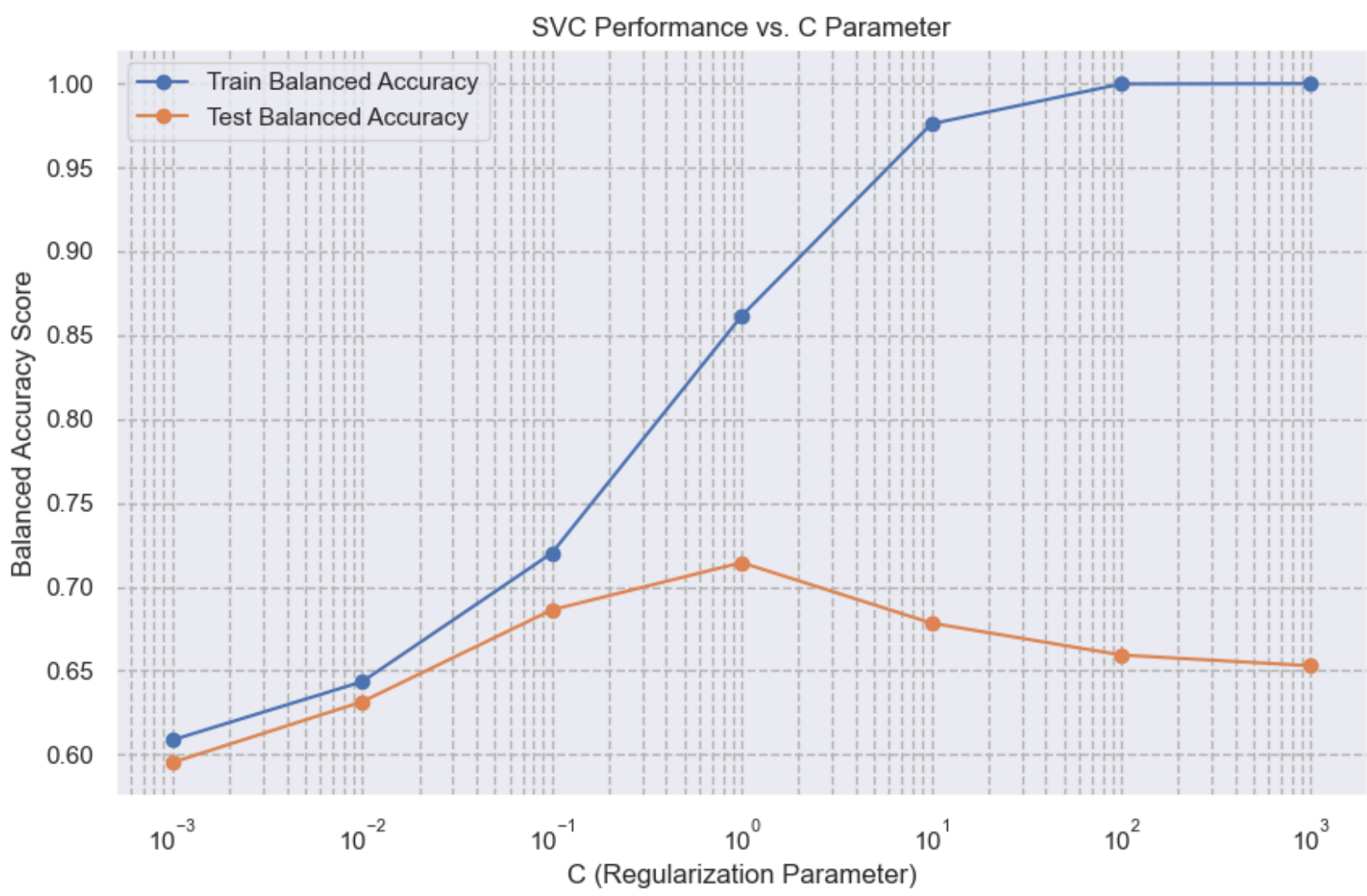

Figure 3.
Scrutinising for overfitting, ensuring that the deviation between train and test accuracy is no higher than 5%.
C: 0.001, Train Balanced Acc: 0.609, Test Balanced Acc: 0.595
C: 0.010, Train Balanced Acc: 0.643, Test Balanced Acc: 0.632
C: 0.100, Train Balanced Acc: 0.720, Test Balanced Acc: 0.686
C: 1.000, Train Balanced Acc: 0.861, Test Balanced Acc: 0.714
C: 10.000, Train Balanced Acc: 0.976, Test Balanced Acc: 0.678
C: 100.000, Train Balanced Acc: 1.000, Test Balanced Acc: 0.659
C: 1000.000, Train Balanced Acc: 1.000, Test Balanced Acc: 0.653

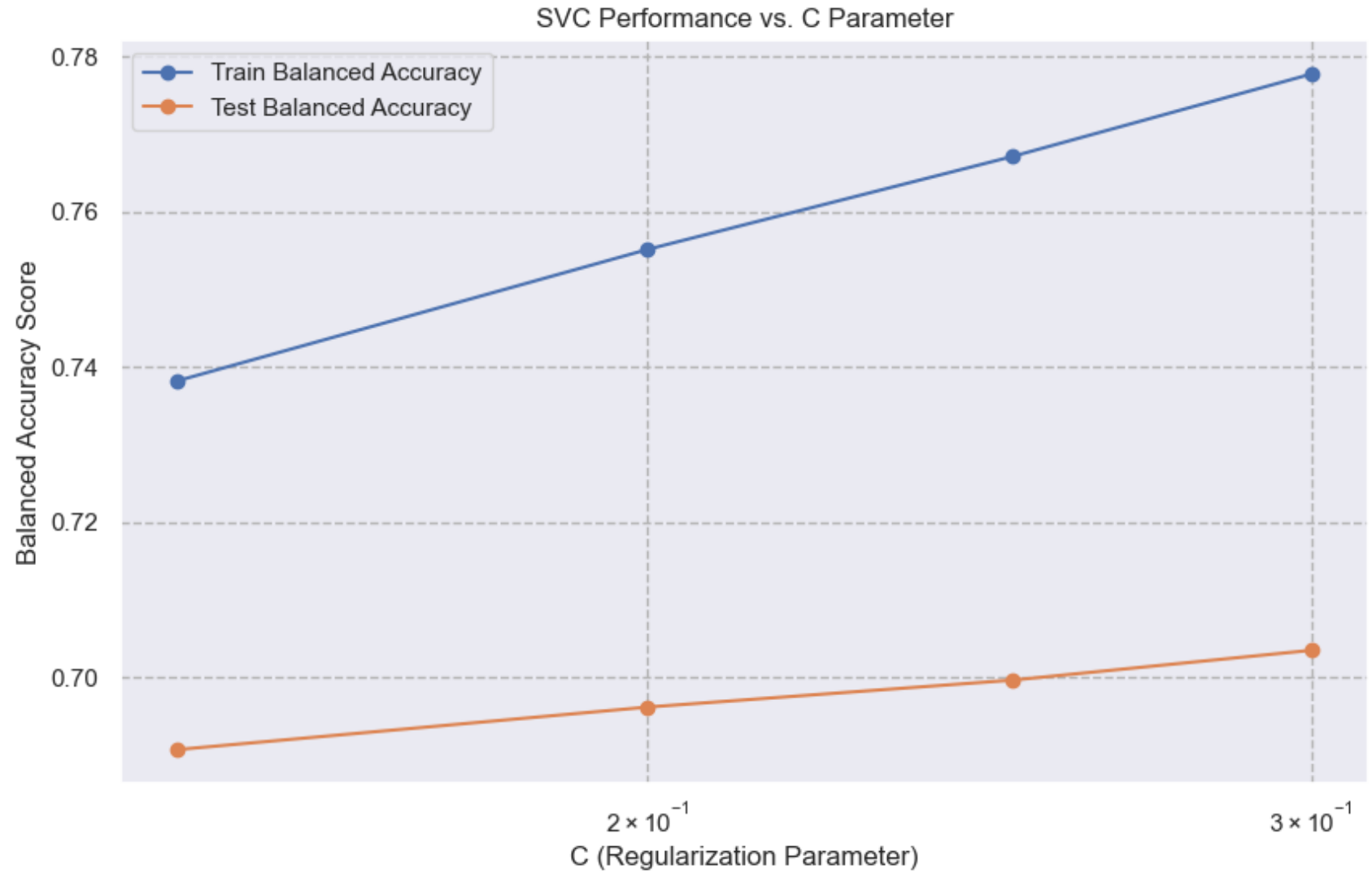

Figure 4.
Scrutinising for overfitting, ensuring that the deviation between train and test accuracy is no higher than 5%.
C: 0.150, Train Balanced Acc: 0.738, Test Balanced Acc: 0.691
C: 0.200, Train Balanced Acc: 0.755, Test Balanced Acc: 0.696
C: 0.250, Train Balanced Acc: 0.767, Test Balanced Acc: 0.700
C: 0.300, Train Balanced Acc: 0.778, Test Balanced Acc: 0.703

Hyperparameter optimisation was conducted via Optuna over 25 trials, identifying the optimal configuration as an SVC with a linear kernel, gamma="auto", shrinking=False, tol=2.88 X $10^{-5}$. The model demonstrated robust generalization, as evidenced by the negligible discrepancy between the training accuracy (71.28%) and the testing accuracy (70.8%), suggesting the absence of significant overfitting. Evaluation of the final model yielded an overall Accuracy of 70.8%, supported by a Precision of 74.3% and A recall of 63.6%. These metrics result in an F1-score of 68.5%, while the AUC-ROC of 0.76 confirms that the classifier maintains an acceptable level of discriminative power for distinguishing between active and inactive target classes.

When using PCA, unsupervised ML learning suggested the number of components to be reduced from 221 to 23. However, the dimensionality reduction with PCA with 23 components did not improve the performance of the ML model. The SVC achieved 64.1% accuracy, 68.6% precision, 51.8 % recall, 59.1% F1 and 64.1% ROC, followed by RFC with 61.9% accuracy, 70.4% precision, 41.1 % recall, 51.9% F1, 61.9% ROC. A dedicated analysis was conducted on the optimisation results observed during the application of PCA. Although unsupervised learning suggested a dimensionality reduction from 221 features to 23 components to capture 50% of the variance, this transformation failed to hold the model performance or improve it. On the contrary, the SVC accuracy decreased to 64.1%, with significant declines across all metrics, including a drop in recall to 51.8%. Several factors likely contributed to the failure of PCA in this specific context. First, the inherent sparsity of the 13C NMR spectroscopic data means that PCA, which relies on linear combinations of all features, may have inadvertently blended critical, low variance signals essential for distinguishing specific antagonistic functional groups. Furthermore, as a linear dimensionality reduction technique, PCA is limited in capturing the likely non-linear and complex relationships between 13 NMR chemical shifts and the pharmacological activity of D1 DAR antagonists. By forcing these relationships into a lower-dimensional linear space, essential structural information was lost. In spectroscopic data, signals accounting for low variance across the dataset may still represent key signatures for active compounds; PCA's prioritisation of high-variance features may have excluded these rare but highly predictive peaks. These findings suggest that for 13C NMR-based machine learning, the full spectroscopic profile (221 features) is necessary to maintain the integrity of the molecular proxy; consequently, the high-dimensional feature set was retained for the final SVC model to ensure maximum predictive reliability.

Upon evaluation of the primary classification metrics, the ML model hyperparameter tuned with Optuna (Model 2) is the mathematically superior option for general purposes, as it leads in Accuracy, Precision, and ROC AUC. Its higher ROC score suggests it has a better fundamental ability to distinguish between classes, while the jump in Precision indicates it is significantly more reliable when it predicts a positive result. However, the choice isn't a total landslide; the ML model based on default hyperparameters and C=0.19 (Model 1**)** maintains a higher Recall, making it the safer bet if your priority is to avoid "missing" cases (False Negatives) at all costs. Since the F1 scores are nearly identical, efficiency is not sacrificed by the selection of either model; therefore, the final decision is dictated by whether a false alarm (Model 2) or a missed detection (Model 1) is assigned greater weight.

This ROC curve of Model 2 demonstrates that the SVC exhibits moderate predictive performance, as evidenced by an AUC of 0.76 (Figure 5). This metric indicates a 76% probability that the model will correctly rank a randomly selected positive instance higher than a negative one. Visually, the curve illustrates the trade-off between the True Positive Rate (Sensitivity) and the False Positive Rate (1 - Specificity) across varying decision thresholds. Because the trajectory inclines toward the upper-left quadrant and remains significantly above the non-discrimination line, the model demonstrates a clear capacity for distinguishing between classes. However, this acceptable level of discrimination suggests that while the classifier is robust for general applications, it maintains a non-negligible false positive rate at higher sensitivity levels, indicating potential for further optimization through hyperparameter refinement or advanced feature engineering.

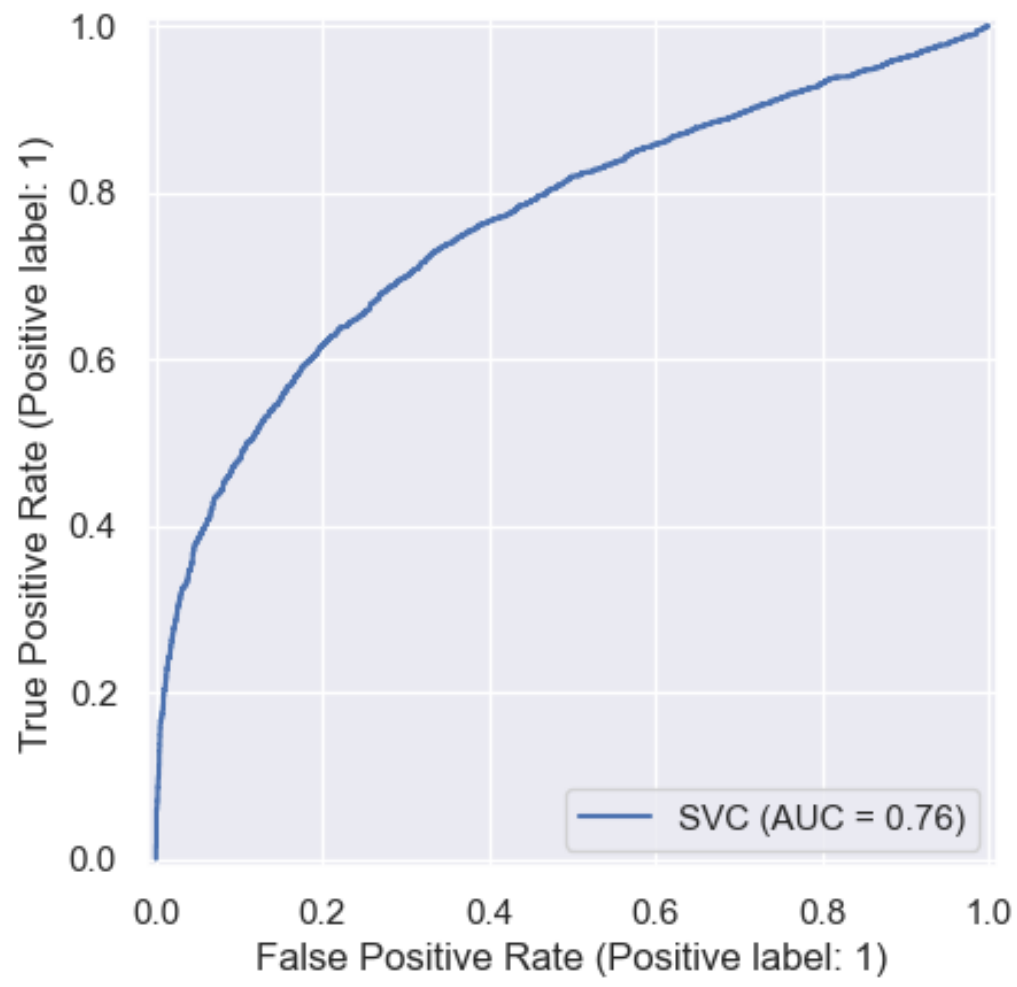

Figure 6. SVC AUC

The confusion matrix of Model 2 (Figure 6) and the MCC of 0.4204 provide a comprehensive snapshot of the classifier's performance at its current decision threshold. With 2,132 True Negatives and 1,738 True Positives, the model demonstrates a strong ability to identify both classes correctly; however, the presence of 995 False Negatives suggests a tendency to miss a notable portion of positive instances in favour of maintaining a lower False Positive count (601). The MCC value of 0.4204 serves as a robust metric for this performance, indicating a moderate-to-substantial positive correlation between the predicted and actual labels. Unlike accuracy, which can be misleading in certain contexts, this MCC score confirms that the model's predictions are significantly better than random chance and possess genuine discriminative power across all four quadrants of the matrix.

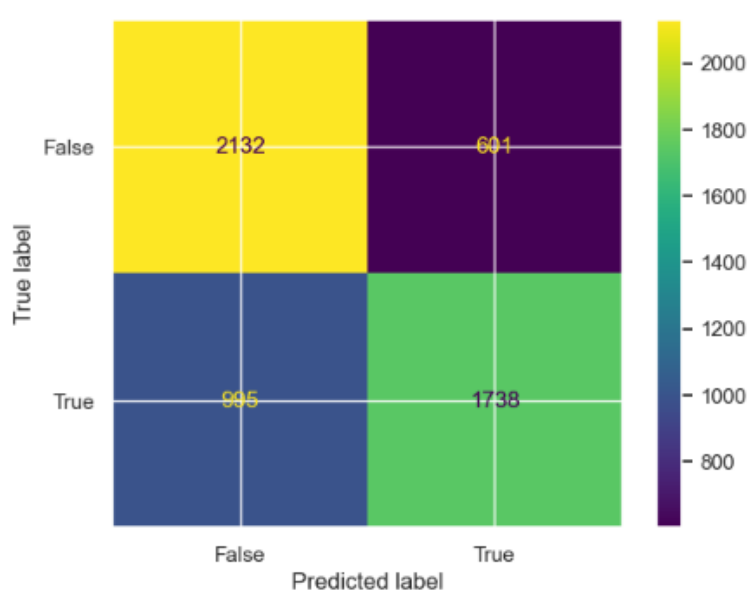

Figure 6. SVC Confusion Matrix

The provided classification report of Model 2 (Table 1) indicates that the model achieves a balanced accuracy of 0.71, reflecting consistent performance across both classes in this perfectly balanced dataset. For the Active (target 1) class, the model prioritizes recall (0.78) over precision (0.68), suggesting it is effective at capturing the majority of active instances but does so at the cost of a higher false-positive rate. Conversely, for the Inactive (target 0) class, the model demonstrates higher precision (0.74) than recall (0.64), meaning its predictions for inactive instances are more reliable, though it misses a larger portion of the actual inactive population. The macro-average F1-score of 0.71 confirms that the classifier maintains a stable and satisfactory level of discriminative power, successfully avoiding a bias toward either label while navigating the inherent trade-offs between sensitivity and precision.

Table 1. Classification report

| | precision | recall | F1-score | support |
|---|---|---|---|---|
| Active (target 1) | 0.68 | 0.78 | 0.73 | 2733 |
| Inactive (target 0) | 0.74 | 0.64 | 0.69 | 2733 |
| | | | | |
| accuracy | | | 0.71 | 5466 |
| macro avg | 0.71 | 0.71 | 0.71 | 5466 |
| Weighted avg | 0.71 | 0.71 | 0.71 | 5466 |

## 3.2. Complementary study`s results

Regarding the CID_SID ML model, which was not a primary object of exploration in the presented study, which is not nano-bio related research and was developed in favour of the researchers interested in human dopamine D1 receptor antagonists, the estimator presented the best was XGBC achieving Accuracy of 80.6%, Precision of 86.4%, Recall of 72.6%, F1-score of 78.8% and ROC of 80.6%, closely followed by GBC. The statistical comparison between XGBC and GBC revealed that the difference in performance is not statistically significant. The observed difference is likely due to random chance. The exploration continued with XBGC whose five-fold cross-validation obtained average of Accuracy of 0.8164±0.007, Precision of 0.8063±0.0112, Recall of 0.7016±0.0092, F1-score of 0.7503±0.0094 and ROC of 0.8589±0.0042. The cross-validation results demonstrate a more robust and reliable model compared to the XGBC single run, primarily due to the high ROC-AUC (0.8589) and the stability indicated by the low standard deviations. While the XGBC single run reports a higher Precision (86.4%) and F1-score (78.8%), these figures lack the statistical backing of multiple iterations and likely reflect an advantageous data split rather than superior generalization. The cross-validation's significantly higher ROC-AUC suggests a much stronger ability to distinguish between classes across various thresholds, and the tight margins (e.g., ±0.007 for Accuracy) provide high confidence that this performance will remain consistent on unseen data. In short, the CV results represent the model's true capability, whereas the single run overestimates specific metrics like Precision at the expense of overall discriminative power.

The hyperparameter tuning for max_depth reveals a clear transition from underfitting to overfitting, with optimal generalisation occurring at max_depth = 4, where the model achieves a peak test accuracy of 81.2% (Figure 7). Applying a 5% threshold for deviation, meaning the model was considered as stable only when the gap between training and testing accuracy is within approximately 0.04 units, the model remains technically acceptable up to max_depth = 5 (where the gap is exactly 0.025). Beyond this point, specifically starting at max_depth = 6 (gap of 0.036) and accelerating rapidly after max_depth = 8 (gap of 0.057), the model exceeds the 5% threshold, signaling that it is becoming too complex and is overfit to the training data. Consequently, while the training accuracy continues to climb toward 90%, the declining test accuracy and widening gap confirm that any depth beyond 5 compromises the model's reliability and predictive power on unseen datasets. At max_depth=4, XGNC XGBC achieved Accuracy of 80.6%, Precision of 87.6%, Recall of 71.2%, F1-score of 78.6% and ROC of 80.6%

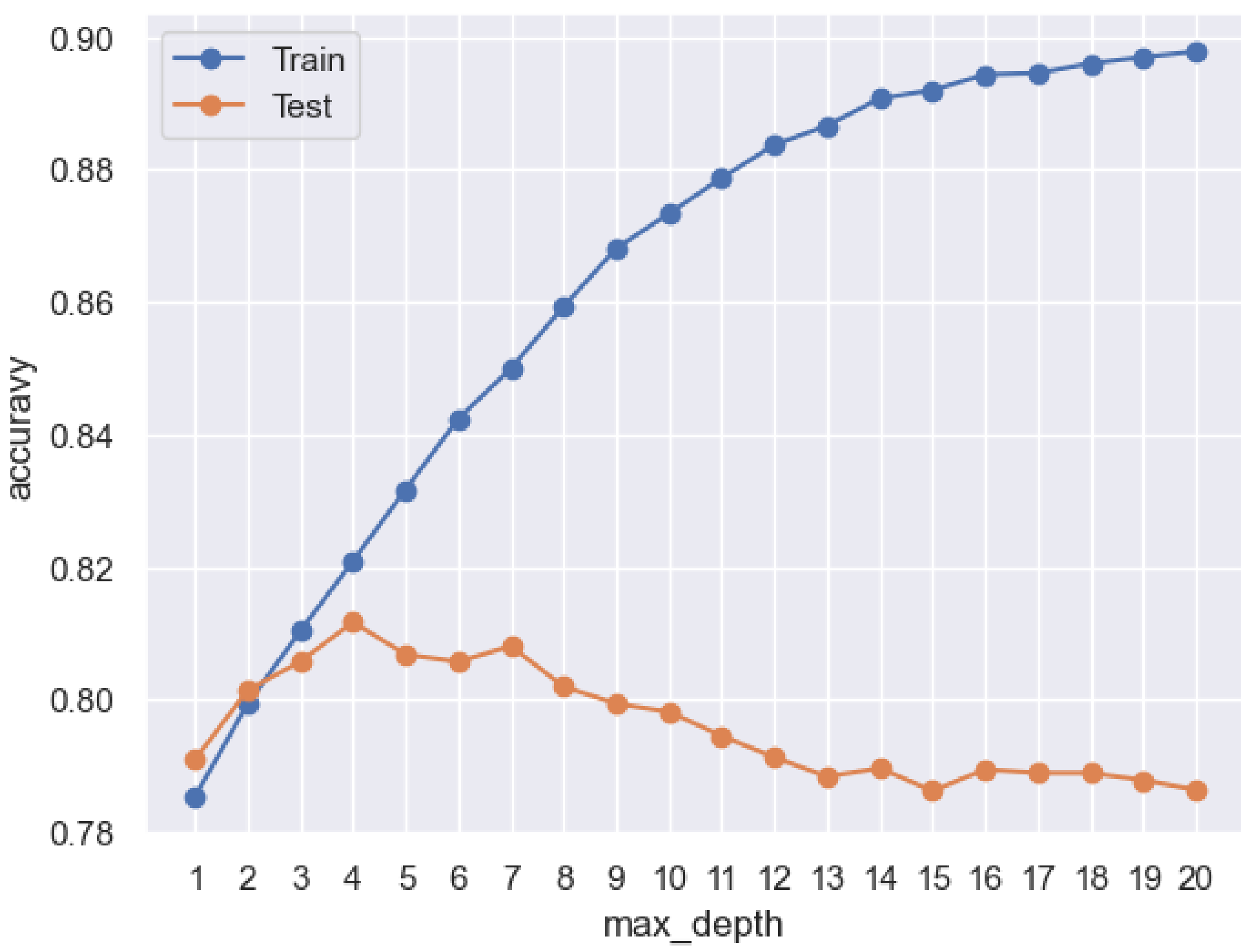

Figure 7.
Impact of tree depth on model generalisation: identifying the overfitting threshold

The ML model was hyperparameter tuned by grid search, and the resulting hyperparameter values were as follows:

(i) colsample_bytree=1.0 (that randomly selects a subset of features for each tree, reducing the correlation between trees in the ensemble and thus preventing overfitting and improving the generalisation)
(ii) learning_rate = 0.3 (determine how much the model`s parameters are adjusted during each iteration of the training process)
(iii) max_depth=4 (limits the number of levels or nodes the model has from the root node to the leaf nodes)
(iv) min_child_weight=1 (control the minimum number of samples required to create a new node in a tree)
(v) n_estimators=300 (denotes the number of decision trees within SVC)
(vi) subsamples=1.0 (It controls the fraction of training data randomly sampled for each tree in the ensemble)

The Grid-search hyperparameter tuned XGBC achieved Accuracy of 81%, Precision of 86.2%, Recall of 73.7%, F1-score of 79.5% and ROC of 81%. The hyperparameter tuning with 25 trials with Optuna suggested max_depth=4, colsample_bytree=0.642865867007244, learning_rate=0.0463091310792728, min_child_weight=4, n_estimators=454, subsample=0.5158921363811282,

gamma=0.8189221399026596, reg_lambda=0.006708296030303976, base_score=0.5) achieving Accuracy of 80.3%, Precision of 88.1%, Recall of 70.1%, F1-score of 78.1% and ROC of 80.3%. Comparing the three variants of XGBC as an ML model with default hyperparameters and max_depth=4 (Model 1), ML model with hyperparameter s tuned by Grid-search (Model 2) and an ML model hyperparameter tuned by Optuna (Model 3), Model 1 being identified as the superior configuration due to its leading Accuracy of 0.806 and an F1-score of 0.786. Although the highest Precision (0.881) was produced by Model 3, this gain was offset by the lowest observed Recall (0.701), suggesting a trade-off that favoured specificity over sensitivity. Conversely, the greatest Recall (0.718) was exhibited by Model 2, though its overall discriminative power, as measured by the ROC-AUC of 0.803, remained slightly below that of the primary candidate. Ultimately, Model 1 was selected for implementation because the most stable and generalized performance across all metrics was maintained by this iteration, ensuring a balanced approach to both false positive and false negative error rates.

The comparative performance of the CID_SID ML model, ML model based on MORGAN2 features and ML model based on RDKit transformed SMILES was analysed to determine the most effective approach for classification (Figure 8). The CID_SID ML model was identified as the superior configuration, as the highest MCC of 0.6225 was achieved by this iteration. This model demonstrated exceptional predictive accuracy, correctly identifying 1,970 True Positives while maintaining the lowest rates of both False Positives (278) and False Negatives (797). In contrast, significantly lower predictive reliability was exhibited by the MORGAN2 and RDKit SMILES models, which yielded MCC scores of 0.3791 and 0.551, respectively. Furthermore, a remarkable advantage in computational efficiency was observed, as a processing time of 2.3544 seconds was recorded for the CID_SID model, outperforming the RDKit SMILES model by a factor of 14. Consequently, the CID_SID approach was selected for final deployment, as the most robust balance between statistical significance and processing speed was consistently maintained by this model.

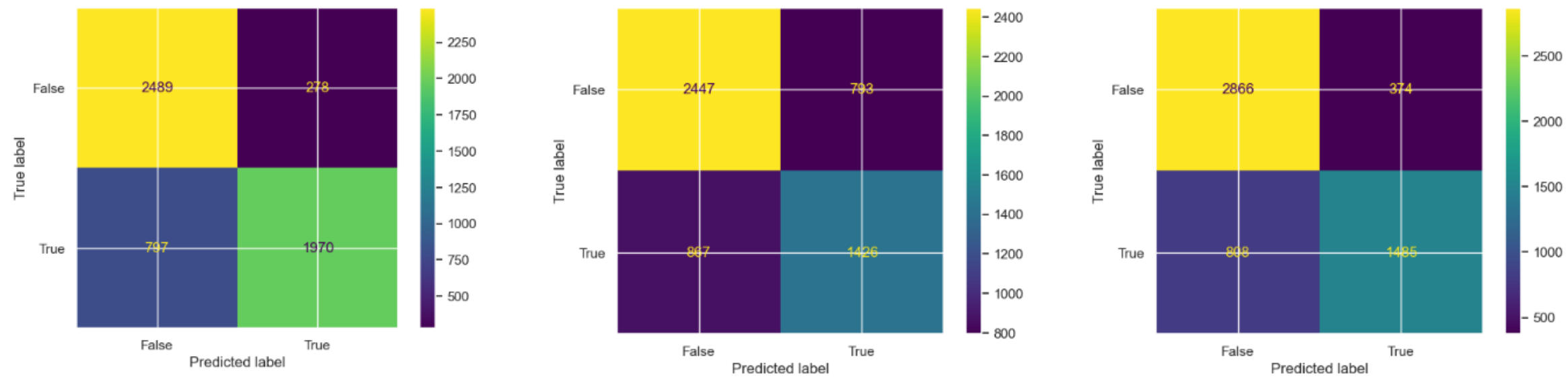

Figure 8.
Comparison of Matrices of XGBC
with CID_SID data (a), MORGAN2 (b) and RDKit SMILES transformed data(c)

Crucially, the presented methodologies ensure the resulting predictions are free of AI hallucination [43, 44]. This is achieved because the ML approach is strictly non-generative, relying exclusively on NMR chemical shifts computationally derived from SMILES structures. By grounding the model in these empirically verifiable spectroscopic features, the integrity and chemical relevance of the predicted results are maintained.

## 4. Discussion

The development of the ML model based on 13C NMR spectroscopy represents a foundational component of the broader Lucid and Universal Bio-applicability Evaluation of Nanoformulation (LUBEN) research initiative. LUBEN is expected to be designed as an ensemble framework, integrating diverse ML architectures and bioassay data to predict the functional impact of NP on conjugated small biomolecules. A primary limitation of the current study is the absence of experimental laboratory validation, resulting from lack of access to specialised facilities. However, the model and the overarching LUBEN framework provide a robust theoretical basis for establishing dynamic portfolios of nanoformulations. This methodology hold the potential to facilitate the exploration of NP behavior and its subsequent influence on biomolecular activity over time and across varying environments, and thus, hold a potential for generation of an NP impact portfolio, collecting information for the NP-biomolecule conjugates would be subjected to various environments, including in vitro, in situ, and organ-on-a-chip systems, that mimic human physiological fluids and organ conditions. By scanning these conjugates via 13C NMR spectroscopy across discrete time windows, the resulting chemical shifts can be utilised as longitudinal descriptors for functional prediction. It should be emphasised here that while the ML model was trained on data not explicitly derived from nanoparticles, its reliance on high-resolution NMR spectral signatures, deconstructed to the atomic level, enables its application as a predictive tool for nanoparticle-influenced systems, thereby bridging the gap in available nanomedicine datasets.

**Model Performance and Real-World Utility**

While the SVC achieved a 71.5% accuracy rate, implying a 28.5% error rate, this performance represents a statistically significant advancement within the context of early-stage virtual screening for nanoparticle-biomolecule interactions. In drug discovery and nanomedicine, the primary objective of ML models is often enrichment rather than absolute prediction; a model with >70% accuracy serves as a potent computational filter capable of narrowing down massive chemical libraries to high-probability subsets. By effectively eliminating roughly 70% of clearly inactive or incompatible compounds, this framework dramatically reduces the 'wet-lab' workload, offering substantial savings in both time and experimental costs. Furthermore, in a field currently hindered by a scarcity of large-scale, labeled experimental datasets for nanoparticle-molecule conjugates, this model provides a robust baseline using in silico 13C NMR data as a molecular proxy. The 71.5% accuracy demonstrates that chemical-shift data alone can capture sufficient functional information to guide researchers toward active antagonists even in the absence of expensive experimental spectra. As indicated by the learning curves (Figure 2), the model's performance has not yet reached a plateau, suggesting that this accuracy serves as a 'floor' for future scaling as more diverse datasets become available.

The predictive capacity of the LUBEN framework is expected to be enhanced by augmenting the existing dataset with Atomic Force Microscopy (AFM) descriptors. Unlike 13C NMR spectroscopy, which is primarily utilised as a proxy for biomolecular environments, AFM is widely recognised and effectively utilised as a cornerstone nanoscience technique for obtaining high-resolution structural and material characterisation of NPs. Consequently, the performance of the ML models is expected to be enhanced by the exploration of both multimodal (combined NMR and AFM) and unimodal (separate NMR or AFM) variants to identify complex functional patterns. The integration of numerical, tabular NMR data with graphical AFM image data into a single, cohesive predictive model is facilitated by Multimodal Machine Learning (also referred to as Multimodal Fusion). This is achieved through the implementation of a specialised neural network architecture wherein separate processing pathways are established for each distinct data modality. These disparate features are then merged at a fusion layer prior to the final prediction, allowing for a more holistic characterisation of the nanoformulation. The feasibility of this approach is supported by the availability of appropriate AFM datasets, such as those published by Carracedo-Cosme et al. [45]

A comparative evaluation of the CID_SID ML model performance was conducted to assess its consistency between the global source study and its specific application to dopamine D1 receptor antagonists (Table 2). While the model maintained a high degree of predictive reliability, a slight reduction in overall effectiveness was observed in the specialized dataset. Specifically, a decrease in Accuracy from 83.5% to 80.6% and a marginal decline in Precision from 89.62% to 87.6% were recorded. The most significant variance was noted in the Recall, which dropped from 75.65% to 71.2%, resulting in a corresponding decrease in the Matthews Correlation Coefficient (MCC) from 0.6795 to 0.6225. Despite these shifts in classification metrics, a notable advantage in computational efficiency was demonstrated, as the processing time for the D1 antagonist task was reduced to 2.3544 seconds compared to the study mean of 3.3079 seconds. Ultimately, it is concluded that while specialized chemical spaces may slightly attenuate the model's sensitivity, the CID_SID configuration remains a robust and high-speed tool for targeted pharmacological research.

Table 2. Comparative performance metrics of the CID_SID ML model: source study vs. Dopamine D1 receptor antagonist dataset

| **Metrics** | **CID_SID ML model mean values from the source study** | **CID_SID ML model values based on dopamine D1 receptor antagonists** |
|---|---|---|
| Accuracy | 83.5% | 80.6% |
| Precision | 89.62% | 87.6% |
| Recall | 75.65% | 71.2% |
| F1-score | 81.93% | 78.6% |
| ROC | 83.53% | 80.8% |
| MCC | 0.67945 | 0.6225 |
| time | 3.3079 seconds | 2.3544 seconds |

The integration of the building units presented in the article for various case studies into the LUBEN or CID_SID ML-based frameworks would assist not only drug development but also risk assessment policies regarding clinical trials. Given that nine out of ten drug candidates fail to reach the approval stage, clinical trials remain a primary focus for time and cost-reduction goals where AI-driven interventions could provide significant benefits [46].

## Conclusion

In this study, a novel computational methodology for predicting the functionality of small biomolecules based on their 13C NMR spectra was successfully developed and demonstrated through the assessment of D1 DAR activity. Although nanoformulations were not included in the primary datasets, it was hypothesized, based on established scientific logic, that the approach remains applicable for predicting the influence of NPs on tethered small biomolecules, provided that the nanoformulations are scanned on physical 13C NMR spectrometers to avoid the high error rates associated with in-silico chemical shift predictions in the presence of NPs. While the spectroscopic data utilized for this proof-of-concept were computationally generated, the framework was specifically designed for direct application with inputs derived from actual 13C NMR spectroscopy. The versatility of the presented framework is emphasized; although validated for D1 DAR antagonists, it is suggested that the model can be extended to predict diverse functionalities for any small biomolecule, provided the data requirements of the ML model are met. Furthermore, the integration of such models into a unified framework, termed LUBEN, is proposed to facilitate the portfolio generation of nanoformulations. It was observed that model performance was positively influenced by increased sample sizes, indicating that further refinements will be realized through dataset expansion, while the development of a supplementary CID_SID ML model marks a significant step toward predicting human D1 DAR antagonist potential using only PubChem identifiers. Ultimately, these ML models are presented as robust predictive tools, though it is acknowledged that laboratory confirmation was not performed due to limited facility access; therefore, experimental validation by the research community is encouraged.

## Author Contributions

MLI, MN, NR, GM and KN conceptualized the project and designed the methodology. MLI and NR wrote the code. MLI, NR and GM processed the data. KN supervised the project. All authors were involved with the writing of the paper.

## Acknowledge

MLI thanks the UWL Vice-Chancellor's Scholarship Scheme for their generous support. We thank NIH for providing access to their PubChem database. The article is dedicated to Luben Ivanov.

## Data and Code Availability Statement

The Python code is available in Jupyter notebooks files on GitHub:
https://github.com/articlesmli/13C_NMR_ML_model_D1.git .

Datasets generated during the study, available on Hugging face:

- 13C NMR spectroscopy data for dopamine D1 receptor antagonists [47]
- Dataset for the CID_SID ML model predicting dopamine D1 receptor antagonists [48]

Raw data used in the study, provided by PubChem:

- AID 2732 Antagonists of Human D 1 Dopamine Receptor: qHTS [19]
- AID 1996 Aqueous Solubility from MLSMR Stock Solutions. *PubChem*. National Institutes of Health [23]

## Conflicts of Interest

The authors declare no conflict of interest.

## Ethics declaration

It is not applicable.

## Funding declaration

It is not applicable.